\documentclass[a4paper,preprint]{emulateapj}
\usepackage{amstext}
\usepackage{apjfonts}
\usepackage{amsmath}

\begin{document}

\title{High Energy Emission from the Double Pulsar system J0737-3039}

\author{Jonathan Granot and Peter M\'esz\'aros\altaffilmark{1}}
\affil{Institute for Advanced Study, Einstein Drive, Princeton, NJ
08540; granot@ias.edu, pim@ias.edu}

\altaffiltext{1}{also Pennsylvania State University, Dpt. of Astronomy \&
Astrophysics, Dpt. of Physics and Center for Gravitational Physics}

\begin{abstract}

We discuss the effects of particle acceleration at the bow shocks
expected in the binary pulsar system J0737-3039, due to the wind
from pulsar A interacting with both the interstellar medium (ISM),
and with the magnetosphere of pulsar B. In this model, we find
that the likeliest source for the X-rays observed by Chandra is
the emission from the shocked wind of pulsar A as it interacts
with the ISM. In this case, for favorable model parameter values,
better statistics might help Chandra marginally resolve the
source. A consequence of the model is a power law high energy 
spectrum extending up to $\lesssim 60\;$keV, at a level of
$\sim 2\times 10^{-13}\;{\rm erg\;cm^{-2}\;s^{-1}}$.

\end{abstract}


\section{Introduction}

The double radio pulsar system J0737-3039 \citep{Lyne04,Kaspi04} is of
great interest as a remarkable laboratory for probing strong field
gravity and magnetospheric interactions.  It has also been detected in
a $10\;$ks Chandra observation \citep{M04}, with an X-ray luminosity
of $L_X \approx 2\times 10^{30}(d/0.5\; {\rm kpc})^2\;{\rm
  erg\;s^{-1}}$ in the $0.2-10\;$keV range (where $d$ is the distance
to the source), and a reported X-ray
photon number index of $\Gamma=2.9\pm 0.4$. The spin-down luminosity
of pulsar A, which is expected to be channeled mainly into its
relativistic wind, is ${\dot E}_A\simeq L_A\simeq 6\times 10^{33}$ erg
s$^{-1}$ \citep{Lyne04,Kaspi04}. Since $L_A\sim 3\times 10^3 L_X$,
only a small fraction of $L_A$ is required in order to produce the
observed X-ray emission.  Since only $77\pm 9$ X-ray photons were
detected, the determination of the spectral slope is difficult, and
might be consistent with a flat $\nu F_\nu$ ($\Gamma \sim 2$), as
expected from shock acceleration. Here, we explore whether particle
acceleration in the bow shocks of the pulsar A relativistic wind can
explain the properties of the X-ray emission. The bow shock on the the
magnetosphere of pulsar B involves only a small fraction of the pulsar
A wind, due to the small solid angle it extends as seen from pulsar A.
Therefore, it must have a very high radiative efficiency in order to
explain the observed X-ray luminosity. On the other hand, the bow
shock on the interstellar medium (ISM) involves most of the pulsar A
wind and thus allows for a significantly smaller and more realistic
radiative efficiency. We evaluate the expected high energy emission
from this shock model, which also predicts emission up to tens of keV.

\section{Emission from the Bow Shock on the ISM}
\label{ISM}

At a sufficiently large distance from the double pulsar system, a bow
shock forms due to the interaction of the wind from pulsar A with the
interstellar medium (ISM).\footnote{The spin-down power of pulsar B is
  $\sim 3\times 10^3$ times smaller than that of pulsar A, so that its
  wind should have a negligible effect on the bow shock with the ISM.}
This situation is similar to that for a millisecond pulsar with a
close low mass binary companion \citep{AT93}, as far a the
interaction between the pulsar wind and the ISM is concerned. The
relative velocity of the center of mass of the binary pulsar with
respect to the ISM is $140.9\pm 6.2\;{\rm km\; s^{-1}}$ on the
plane of the sky \citep{Ransom04}. A velocity component along
our line of sight could lead to a larger total velocity, $v_{\rm
ext}=200 v_{200}\;{\rm km\; s^{-1}}$ with $v_{200}\gtrsim 1$. The
head of the bow shock is at a distance $R$ from pulsar A where the
kinetic pressure of the wind balances the ram pressure of the
ambient medium, $\rho_{\rm ext}v_{\rm ext}^2$,
\begin{equation}\label{R_2}
R=\sqrt{\frac{L_A}{4\pi\rho_{\rm ext}v_{\rm ext}^2 c}}= 4.9\times
10^{15}\,n_0^{-1/2}v_{200}^{-1}\;{\rm cm}\ ,
\end{equation}
where $\rho_{\rm ext}=n_{\rm ext}m_p$ and $n_{\rm ext}=n_0\;{\rm
cm^{-3}}$ are the ambient mass density and number density,
respectively.

Pulsar winds are thought to have a pair plasma composition, perhaps
with ions in restricted latitude sectors, and a high asymptotic bulk
Lorentz factor (perhaps as high as $\sim 10^4 - 10^6$ in the Crab
Nebula and other young pulsar wind nebulae). For simplicity we assume
a pure $e^\pm$ pair plasma which holds a fraction $\epsilon_e\approx
1$ of the internal energy behind the shock.  We use a fiducial value
of $\gamma_w=10^5\gamma_{w,5}$ for the wind Lorentz factor just before
the shock, however our main results are rather insensitive to the
exact value of $\gamma_w$. We assume $\gamma_w\gg 1$ throughout this
work.

The ratio $\sigma$ of Poynting flux to kinetic energy in the wind is
believed to be $\sigma\gg 1$ at very small radii, while low values of
$\sigma\ll 1$ at large radii are inferred from observations
\citep[e.g. $\sigma\sim 3\times 10^{-3}$ for the Crab,][]{GA94,SA04}.
It is hard to estimate the value of $\sigma$ at intermediate radii,
which are relevant for our purposes. For the bow shock with the ISM
which is at a relatively large radius, one might expect
$\sigma\lesssim 1$. The shock jump conditions imply that the fraction
$\epsilon_B$ of the internal energy behind the shock in the magnetic
field is $\epsilon_B\sim\sigma$. However, amplification of the
magnetic field in the shock itself could produce $\epsilon_B\sim 1$
even if $\sigma\ll 1$. Conversely, for $\sigma>1$ magnetic dissipation
behind the shock might decrease the value of $\epsilon_B$ and make it
close to unity. Therefore, we assume $\epsilon_B\sim 1$, and to zeroth
order we neglect the effect of the magnetic field on the shock jump
conditions.

In order to estimate the emission from the shocked wind, we will
use the values of the hydrodynamical quantities at the head of the
bow shock. To first order we neglect the orbital motion of pulsar
A. The proper number density in the wind, as a function of the
distance $r$ from pulsar A, is $n_w=L_A/4\pi r^2m_e
c^3\gamma_w^2$. The shock jump conditions at
the head of the bow shock imply that the shocked pulsar A wind just
behind the shock moves away from the shock at $\beta=1/3$, and has
a proper energy density $e_{\rm int}=L_A/2\pi R^2c\approx
1.3\times 10^{-9}n_0 v_{200}^2\;{\rm erg\;cm^{-3}}$ and a proper
number density $n=2^{3/2}\gamma_wn_w=e_{\rm
int}/(m_ec^2\gamma_w/\sqrt{2})=2.3\times 10^{-8}n_0
v_{200}^2\gamma_{w,5}^{-1}\;{\rm cm^{-3}}$. This implies a magnetic
field of $B=1.8\times 10^{-4}n_0^{1/2}v_{200}\epsilon_B^{1/2}\;$G
(in the fluid rest frame). The $e^\pm$ pairs are assumed to be
accelerated by the shock into a power law energy distribution 
$dn/d\gamma_e\propto\gamma_e^{-p}$, with
$\gamma_m<\gamma_e<\gamma_{\rm max}$. Observations of synchrotron
emission from electrons accelerated in relativistic collisionless
shocks typically imply $p\sim 2-3$. The average random Lorentz
factor of the shocked electrons is
$\langle\gamma_e\rangle=\gamma_w/\sqrt{2}$, and the minimal
Lorentz factor is given by\footnote{More generally, this
expression should be multiplied by a factor of
$(1+\rho_p/\rho_e)$, which can be as high as $m_p/m_e$ in the
limit of a proton-electron plasma. Also, the factor of
$\frac{p-2}{p-1}$ is valid for $p>2$, while for $p=2$ it should be
replaced by $1/\ln(\gamma_{\rm max}/\gamma_m)$ so that
$g=3/\ln(\gamma_{\rm max}/\gamma_m)$.}
\begin{equation}\label{gamma_m}
\gamma_m=\frac{p-2}{p-1}\langle\gamma_e\rangle=
\frac{g\epsilon_e\gamma_w}{3\sqrt{2}}=2.4\times 10^4
g\,\epsilon_e\gamma_{w,5}\ ,
\end{equation}
where $g\equiv 3(p-2)/(p-1)$ equals $1$ for $p=2.5$. The maximal
Lorentz factor, from the requirement that the Larmor radius 
$R_L=\gamma_e m_e c^2/eB$ does not exceed the width $\eta R$ 
of the layer of shocked fluid, is
\begin{equation}\label{gamma_max1}
\gamma_{\rm max,1}=\frac{eB\eta R}{m_ec^2}=1.2\times
10^7\epsilon_B^{1/2}(\eta/0.3)\ .
\end{equation}
Here, the value of $\eta$ can be estimated by equating the
particle injection rate into the hemisphere
containing the head of the bow shock ($\theta\leq 90^\circ$),
$\dot{N}/2=L/2\gamma_w m_e c^2$, to the flow of shocked particles
behind the shock outside of this hemisphere, $2\pi\eta R^2nu$,
where $n=2^{3/2}\gamma_w n_w=\dot{N}/\sqrt{2}\pi R^2 c$ and
$u=\gamma\beta$ are the proper density and 4-velocity (in the
direction perpendicular to the shock) of the shocked wind at
$\theta=90^\circ$. This gives $\eta\approx 1/(2^{3/2}u)$,
so that $\eta<1$ implies $\beta>1/3$. At $\theta=90^\circ$ we
expect $\beta\gtrsim c_s/c\approx 3^{-1/2}$ and $u\gtrsim
2^{-1/2}$ so that $\eta\lesssim 1/2$. On the other hand $\eta<0.1$
requires $u>5/\sqrt{2}\approx 3.5$ which begins to be highly super
sonic, and is therefore not very reasonable. Hence we expect
$0.1\lesssim\eta\lesssim 0.5$ and use a fiducial value of
$\eta=0.3$.

The dominant emission mechanism is synchrotron radiation, and
inverse Compton scattering can be neglected. The Lorentz factor of
an electron which cools on the dynamical time, $t_{\rm dyn}\sim
R/(c/3)=4.9\times 10^5 n_0^{-1/2}v_{200}^{-1}\;$s, is given by
\begin{equation}\label{gamma_c}
\gamma_c = \frac{6\pi m_e c}{\sigma_T B^2 t_{\rm dyn}} =
\frac{4.7\times 10^{10}}{\epsilon_B n_0^{1/2}v_{200}}\ .
\end{equation}
The synchrotron spectral break frequencies 
corresponding to $\gamma_m$, $\gamma_c$ and
$\gamma_{\rm max}$ are
\begin{eqnarray}\label{nu_m2}
\nu_m &=& 3.4\times 10^{11}\,g^2\epsilon_B^{1/2}\epsilon_e^2
n_0^{1/2}v_{200}\gamma_{w,5}^2\;{\rm Hz}\ ,
\\ \label{nu_c2}
h\nu_c &=& 5.5\,\epsilon_B^{-3/2} n_0^{-1/2}v_{200}^{-1}\;{\rm
GeV}\ ,
\\ \label{nu_max2}
h\nu_{\rm max} &=&
62\,n_0^{1/2}v_{200}\epsilon_B^{3/2}(\eta/0.3)^2\;{\rm keV}\ .
\end{eqnarray}
We have $\nu F_\nu\propto\nu^{4/3}$ for $\nu<\nu_m$, $\nu
F_\nu\propto\nu^{(3-p)/2}$ for $\nu_m<\nu<\min(\nu_c,\nu_{\rm
max})$, and if $\nu_{\rm max}>\nu_c$ (which is relevant for the
next section) we have $\nu F_\nu\propto\nu^{(2-p)/2}$ for
$\nu_c<\nu<\nu_{\rm max}$.

Since $\gamma_c>\gamma_{\rm max}$, all electrons radiate only a
small fraction of their energy. The fraction of energy radiated by
an electron is $\sim\min(1,t_{\rm
dyn}/t_c)=\min(1,\gamma_e/\gamma_c)$, where $t_c=6\pi m_e
c/\sigma_T B^2\gamma_e$ is the synchrotron cooling time. Averaging
over the power law electron energy distribution, we obtain
the total fraction $\epsilon_{\rm rad}$ of energy in electrons
that is radiated away. For $\gamma_c<\gamma_m$ (fast cooling),
$\epsilon_{\rm rad}\approx 1$, since all electrons cool
significantly within $t_{\rm dyn}$. For
$\gamma_m<\gamma_c<\gamma_{\rm max}$,
\begin{equation}\label{epsilon_rad}
\epsilon_{\rm rad}\approx \left\{\begin{matrix}1 & p<2 \cr\cr
\left[1+\ln(\gamma_{\rm max}/\gamma_c)\right]/\ln(\gamma_{\rm
max}/\gamma_m) & p=2 \cr\cr (3-p)^{-1}(\gamma_m/\gamma_c)^{p-2} &
2<p<3
 \end{matrix}\right.\ ,
\end{equation}
while for $\gamma_c>\gamma_{\rm max}$ we have $\epsilon_{\rm
rad}(\gamma_c>\gamma_{\rm max})\sim(\gamma_{\rm
max}/\gamma_c)\epsilon_{\rm rad}(\gamma_m<\gamma_c<\gamma_{\rm
max})$, or
\begin{equation}\label{epsilon_rad2}
\epsilon_{\rm rad}\approx\frac{\gamma_{\rm max}}{\gamma_c}\times
\left\{\begin{matrix}\frac{p-2}{3-p} & p<2 \cr\cr
1/\ln(\gamma_{\rm max}/\gamma_m) & p=2 \cr\cr
\frac{p-2}{3-p}(\gamma_m/\gamma_{\rm max})^{p-2} & 2<p<3
 \end{matrix}\right.\ .
\end{equation}
For our fiducial parameters and $p\approx 2$ we have $\epsilon_{\rm
  rad}\approx 4.3\times
10^{-4}n_0^{1/2}v_{200}\epsilon_B^{3/2}(\eta/0.3)$. Most of the
radiated energy will be emitted near $\nu_{\rm max}$ at tens of keV.
The fraction $f_X$ of the radiated energy in the $0.2-10\;$keV Chandra
range is approximately given by the ratio of the average $\nu F_\nu$
value in the Chandra range (equal to the $\nu F_\nu$ value at some
frequency $\nu_X$ within that range) to the peak $\nu F_\nu$ value. In
our case, $f_X\sim (\nu_X/\nu_{\rm max})^{(3-p)/2}\approx
0.27n_0^{-1/4}v_{200}^{-1/2}\epsilon_B^{-3/4}(\eta/0.3)^{-1}$ (this
expression holds for $\nu_X<\nu_{\rm max}<\nu_c$).  Therefore, the
ratio of the expected X-ray luminosity $L_X=f_X\epsilon_e\epsilon_{\rm
  rad}L_A\approx 7\times
10^{29}n_0^{1/4}v_{200}^{1/2}\epsilon_B^{3/4}\;{\rm erg\; s^{-1}}$ in
the Chandra range, to the observed $L_X^{\rm obs} \approx 2\times
10^{30}(d/0.5\;{\rm kpc})^2\;{\rm erg\;s^{-1}}$ is $\sim 0.35
n_0^{1/4}v_{200}^{1/2}\epsilon_B^{3/4}(d/0.5\;{\rm kpc})^{-2}$. This
ratio would be unity, e.g., for $n_0\sim 60$ with $v_{200}\sim 1$, or
for $n_0\sim 10$ and $v_{200}\sim 2.5$.  If $d\approx 1\;$kpc instead
of $\approx 0.5\;$kpc, then we would need $n_0\sim 10^3$ and
$v_{200}\sim 4$, which are less likely.  Conversely, the constraint is
easier to satisfy if $d<0.5\;$kpc.  According to this interpretation
the high energy emission should peak at tens of keV (near $\nu_{\rm
  max}$ that is given in Eq.  \ref{nu_max2}) with a flux of $\sim
L_X^{\rm obs}/f_X 4\pi d^2 \sim 2.5\times
10^{-13}n_0^{1/4}v_{200}^{1/2}\epsilon_B^{3/4}(\eta/0.3)\;{\rm
  erg\;cm^{-2}\;s^{-1}}$.

Another contribution to the X-ray luminosity might be expected from
the shocked ISM in the bow shock. The energy injection rate
is\footnote{Comparing the energy injection rate per unit area into the
  wind termination shock, $L_A/4\pi R^2$, and into the bow shock going
  into the ISM, $(1/2) \rho_{\rm ext}v_{\rm ext}^3$, and balancing the
  two ram pressures, $L_A/4\pi R^2c$ and $\rho_{\rm ext}v_{\rm
    ext}^2$, respectively.}  $\sim(v_{\rm ext}/c)L_A\sim 10^{-3}L_A$,
which is of the order of the observed X-ray luminosity, and perhaps
larger by a factor of a few. This could account for the observed X-ray
luminosity, provided that $f_X\epsilon_e\epsilon_{\rm rad}\gtrsim
0.1-0.3$. Here, the dynamical time is $t_{\rm dyn}\sim R/v_{\rm ext}$
and $h\nu_c = 22\,\epsilon_B^{-3/2} n_0^{-1/2}v_{200}\;$keV,
while $\nu_m$ is very low (in fact $\gamma_m\sim 1$), and the
expression for $\nu_{\rm max}$ is the same as in Eq. \ref{nu_max2}
with the difference that here $\eta R$ is the width of the shocked ISM
layer (instead of the shocked pulsar wind), and that $\epsilon_B$
might be different (probably somewhat smaller) in the shocked ISM. One
might expect instabilities near the contact discontinuity between the
shocked wind and the shocked ISM, both of the Rayleigh-Taylor and
Kelvin-Helmholtz types, which could bring the magnetic field in the
shocked ISM close to equipartition. For $v_{200}\sim 1$ and
$n_0\gtrsim 10$, $\nu F_\nu$ peaks in the Chandra range, so that we
can have $f_X\sim 1$. Since the shock going into the ISM is Newtonian,
one expects $p\approx 2$, as in supernova remnants (SNRs). For
$n_0\sim 60$ this would imply $\epsilon_{\rm rad}\approx 0.2$. From
modeling of collisionless shocks in SNRs, which propagate into a
similar ISM with similar shock velocities, a typical value of
$\epsilon_e\sim 0.1$ might be adopted.  The resulting value of
$L_X\sim(v_{\rm ext}/c)f_X\epsilon_e \epsilon_{\rm rad}L_A\sim 2\times
10^{-5}L_A\sim 10^{29}\; v_{300}(\epsilon_e/0.1)(f_X\epsilon_{\rm
  rad}/0.2)(d/0.5\;{\rm kpc})^2 {\rm erg\; s^{-1}}$ is only $\sim 0.05
L_X^{\rm obs}$. Thus, this emission component may not easily account
by itself for the Chandra observation (unless $\epsilon_e\sim 1$),
although it can contribute to that from the shocked wind of pulsar A.

We note that Eq. \ref{R_2} implies that the angular distance between
the double pulsar system and the head of the ISM bow shock is
$\theta_{bs}=0.65(d/0.5\;{\rm
  kpc})^{-1}n_0^{-1/2}v_{200}^{-1}\;$arcsec, and the relatively bright
emission from the bow shock could extend over an angular scale a few
times larger than this value. This angular scale may be resolved with
Chandra, with longer integration times, even though in the 10 ks
Chandra detection it was reported as a point source \citep{M04}. If
resolved, one might constrain the source angular size to $\lesssim
1\;$arcsec. However, we note that the observed X-ray emission is best
explained from the bow shock with the ISM if
$n_0^{-1/2}v_{200}^{-1}\sim 0.35 \epsilon_B^{3/4}(d/0.5\;{\rm
  kpc})^{-2}$, which in turn implies $\theta_{bs}\approx
0.23\epsilon_B^{3/4}(d/0.5\;{\rm kpc})^{-3}\;$arcsec, that may be
difficult to resolve with Chandra unless $d\lesssim 0.5\;$kpc. On the
other hand, this suggests that $d\gtrsim 0.3\;$kpc, as otherwise the
source should have already been resolved by Chandra, despite the poor
photon statistics in the current observation.

The emission from the bow shock with the ISM is not expected to show
significant modulation at the spin period of pulsar A, $P_A=22.7\;$ms,
or at the orbital period, $P_{\rm orb}=2.45\;$hr.  The former is
because $P_A$ is $\sim 6-7$ orders of magnitude smaller than
$R/c$.\footnote{This has two effects.  First, any variability in the
  wind with the period $P_A$ will be strongly smoothed out by the time
  it reaches the bow shock. Second, the distance of the bow shock from
  pulsar A varies, with $\Delta R\sim R$, so that the phase of the
  pulsar A wind that impinges upon it at any given time changes by
  $\sim 10^6-10^7$ periods. Since the same holds for the observed
  emission, it significantly averages out a possible modulation with a
  period of $P_A$, even if it exists in the local emission from a
  given location along the bow shock.} The latter is because the
orbital velocity of pulsar A is $v_{\rm orb}\approx 300\;{\rm km\;
  s^{-1}}\ll c$, and the distance between pulsars A and B is
$R_{AB}=8.8\times 10^{10}\;{\rm cm}\ll R$.\footnote{The orbital motion
  of pulsar A affects the bow shock with the ISM mainly in two ways.
  First, the distance between pulsar A and the head of the bow shock
  changes by $\sim \pm R_{AB}/2$, changing the ram pressure by $\Delta
  p/p\sim 2R_{AB}/R\sim 10^{-5}-10^{-4}$. Second, the wind is highly
  relativistic and behaves as radiation, so that its intensity in the
  bow shock rest frame scales as the fourth power of the Doppler
  factor $\delta\approx 1+\beta_{\rm orb}\cos\theta$, and will change
  in the range $(1\pm\beta_{\rm orb})^4$, resulting in a relative
  amplitude of $\approx 8\beta_{\rm orb}\approx 0.8\%$.  Since
  $(R/c)/P_{\rm orb}\approx 18n_0^{-1/2}v_{200}^{-1}$, some additional
  averaging can occur due to the different phase of this modulation
  over the different parts of the bow shock, although this effect is
  not very large for our most promising model for which $R$ is smaller
  by a factor of $\sim 8$ compared to its fiducial value in Eq.
  \ref{R_2}.}

\section{Emission from the Bow Shock Around Pulsar B}
\label{pulsar_B}

Balancing the ram-pressure of the pulsar A wind with the magnetic
pressure of pulsar B, assuming a predominantly dipole magnetic field,
and a surface field strength of $B_*=1.2\times 10^{12}\;$G
\citep{Lyne04} the distance of the head of the bow shock measured from
pulsar B is $R_{bs}\approx 6\times 10^9\;$cm. This is $\approx 0.07$
of the separation between the two pulsars, $R_{AB}=8.8\times
10^{10}\;$cm \citep{Lyne04}. Therefore, as seen from pulsar A, the
fraction of the total solid angle subtended by the bow shock is
$\Omega/4\pi=C \pi (R_{bs}/R_{AB})^2/4\pi\approx 10^{-3}C$, where
$C\sim\;$a few, its value depending on the exact shape of the bow
shock. Thus, producing the X-rays in the shocked wind of pulsar A in
the bow shock occurring near pulsar B would require an efficiency of
$\sim 0.3/C$, in order to account for the Chandra observation.

\citet{Lyutikov04} calculated the asymptotic opening angle of the bow
shock, and finds it to be $\theta\sim 0.11-0.13\;$rad for the value of
$B_*$ from \citet{Lyne04}. This gives $\Omega/4\pi \sim (3-4.2)\times
10^{-3}$, which is in agreement with our estimate here, and provides
an independent cross calibration for our parameter $C$, namely $C\sim
2.6-3.6$. 

As in \S \ref{ISM}, the values of the hydrodynamical quantities at the
head of the bow shock are used in order to estimate the emission from
the shocked wind. To zeroth order, we neglect the orbital motion of
the two pulsars, and their spins. The bow shock itself is at rest in
the lab frame, in our approximation.\footnote{We ignore the slower
  binary period timescale, which would cause inertial effects,
  centrifugal and Coriolis, etc., as well as the possible time
  variability due to the rotation of the pulsar B magnetosphere.} The
expressions for the hydrodynamical quantities are similar to those in
\S \ref{ISM}, just that here the distance of the head of the bow shock
from pulsar A, $R=R_{AB}-R_{bs}\approx 8.2\times 10^{10}\;$cm, is
$\sim 10^5$ times smaller. Therefore, we have $e_{\rm int}=4.7\;{\rm
  erg\;cm^{-3}}$, $n=82\gamma_{w,5}^{-1}\;{\rm cm^{-3}}$ and
$B=11\epsilon_B^{1/2}\;$G. While the dynamical time in this case is
much shorter, $t_{\rm dyn}\sim R_{bs}/(c/3)=0.6\;$s, the synchrotron
cooling time $t_c\propto R^2$ is smaller by an even larger factor, so
that $\gamma_c\lesssim\gamma_{\rm max}$. Here $\gamma_{\rm max}$ is
also constrained by radiative losses. This limit may be obtained by
equating the cooling time $t_c$ to the acceleration time, $t_{\rm
  acc}=A(2\pi m_e c\gamma_e/eB)$ where $A\gtrsim 1$, $\gamma_{\rm
  max,2} = (3e/A\sigma_T B)^{1/2} = 1.4\times 10^7
A^{-1/2}\epsilon_B^{-1/4}$.
The limit discussed in \S \ref{ISM} now reads $\gamma_{\rm
max,1}=1.2\times 10^7\epsilon_B^{1/2}(\eta/0.3)$, and we have
$\gamma_{\rm max}=\min(\gamma_{\rm max,1},\gamma_{\rm max,2})$.

Since the bow shock around pulsar B is much more compact than the bow
shock on the ISM, one might expect inverse Compton scattering of the
synchrotron photons to be more important in this case, and therefore
we check this. The Compton y-parameter is given by
$Y\sim\tau_T\gamma_m^{p-1}\gamma_c^{3-p}$ for $2<p<3$,
$Y\sim\tau_T\gamma_m\gamma_c[1+\ln(\gamma_{\rm max}/\gamma_c)]$ for
$p=2$, and $Y\sim\tau_T\gamma_m^{p-1}\gamma_c\gamma_{\rm max}^{2-p}$
for $p<2$ \citep{PK00}. We expect $p\approx 2$, for which $Y(1+Y)\sim
10^{-2}(\eta/0.3)\epsilon_e/\epsilon_B$. For our values of
$\epsilon_e\sim\epsilon_B\sim 1$, this gives $Y\sim 10^{-2}$, so that
inverse Compton scattering is not very important, and is neglected in
our treatment.

We have $\gamma_c=1.1\times 10^7\epsilon_B^{-1}$ and $\gamma_m$ is
still given by Eq. \ref{gamma_m}. The corresponding synchrotron
frequencies are 
\begin{eqnarray}\label{nu_m}
h\nu_m &=& 86\,g^2\epsilon_B^{1/2}\epsilon_e^2\gamma_{w,5}^2\;{\rm eV}\ ,
\\ \label{nu_c}
h\nu_c &=& 17\,(1+Y)^{-2}\epsilon_B^{-3/2}\;{\rm MeV}\ ,
\\ \label{nu_max}
h\nu_{\rm max} &=&
\min[30\,A^{-1}(1+Y)^{-1},20\,\epsilon_B^{3/2}(\eta/0.3)^2]\;{\rm MeV}\ .
\end{eqnarray}
The X-ray luminosity is then $L_X=f_X\epsilon_e\epsilon_{\rm rad}
(\Omega/4\pi)L_A$. For $p\approx 2$ typically $\epsilon_{\rm rad}\sim
0.2$, and $f_X\sim(\nu_X/\nu_c)^{(3-p)/2}\sim 0.02\,\epsilon_B^{3/4}$.
Altogether, and assuming $\Omega/4\pi\sim 4\times 10^{-3}$ ($C\sim
4$), we have $L_X\sim 10^{29}\epsilon_B^{3/4}\;{\rm erg\;s^{-1}}$,
which is a factor of $\sim 20\,\epsilon_B^{-3/4}$ smaller than the
observed value.

It might still be possible to increase $L_X$ if somehow $\nu_c$ could
be lowered, since this would significantly increase $f_X$, and also
somewhat increase $\epsilon_{\rm rad}$. This could potentially be
achieved if $t_{\rm dyn}$ or the magnetic field experienced by the
shocked electrons are increased (as for $p\approx 2$,
$f_X\propto\nu_c^{-1/2}\propto t_{\rm dyn}B^{3/2}$).  This might
happen if a reasonable fraction\footnote{In the bow shock of the solar
  wind around the Earth only $\sim 10^{-3}$ of the wind particles get
  captured by the Earth's magnetic field.  However, the situation
  there is different in several respects from our case. For example,
  the Earth's magnetic field is nearly aligned with its rotational
  axis, while the solar wind is Newtonian ($\sim 400\;{\rm
    km\;s^{-1}}$) with relatively low magnetization and includes
  protons and electrons in roughly equal numbers. Therefore this
  fraction might be larger in our case, and could possibly be
  sufficiently large for our purposes, although this is uncertain.} of
the shocked wind becomes associated with closed magnetic field lines
of pulsar B for one or more rotational periods of B (where $P_B$ is
$\sim 4.6$ times larger than the estimate we used for $t_{\rm dyn}$,
i.e.  $R_{bs}/(c/3)\approx 0.6\;$s). In this case, this material will
also pass through regions of higher magnetic field strength. This
could be the case if, e.g., interchange instabilities cause mixing of
the two fluids across the contact discontinuity.

One might expect a modulation in the emission with the orbital period
due to the change in the line of sight w.r.t. the bow shock
\citep{AT93}.  The shocked wind is expected to move away from the head
of the bow shock with a mildly relativistic velocity, in a direction
roughly parallel to the bow shock \citep{Lyutikov04}.  This might
cause a mild relativistic beaming of the radiation emitted by the
shocked plasma, resulting in a mild modulation \citep[by a factor
$\lesssim 2$,][]{AT93} of the observed emission as a function of the
orbital phase. Another possible source of modulation with the orbital
period may arise if the luminosity of the pulsar A wind depends on the
angle from its rotational axis \citep{D04}. In this case the wind
luminosity in the direction of pulsar B will vary with a period
$P_{\rm orb}$.  The duration of the Chandra observation, $10^4\;$s, is
close to the orbital period $P_{\rm orb}=2.45\;$hr, and it showed no
evidence for variability \citep{M04}. However, the small number of
photons ($77\pm 9$) does not allow to place a strong limit on a
possible modulation with the orbital period, which might still have an
amplitude of $\lesssim 50\%$.

The rotation of pulsar B, assuming some misalignment of its magnetic
pole relative to its spin axis (as expected from the detection of its
pulses), would cause a periodic change in $\Omega$ (with a periodicity
equal to the spin period $P_B=2.77\;$s), with an amplitude which is
typically of order unity \citep{Lyutikov04}. The distance of the bow
shock from pulsar A hardly changes, and therefore the values of the
thermodynamic quantities in the shocked wind and the resulting values
of $f_X$ and $\epsilon_{\rm rad}$ should vary with a smaller
amplitude. Thus, the modulation in $L_X$ is expected to largely follow
that in $\Omega$, and have a similar amplitude (typically of order
unity).

\section{Discussion}

Particle acceleration is expected in the binary pulsar system
J0737-3039 both from the bow shock of the pulsar A wind as it
interacts with the interstellar medium (ISM), and the bow shock of
the wind of pulsar A interacting with the magnetosphere of pulsar B.
The rotational energy loss rate, the systemic velocity and the orbital
separation determine the effective angles subtended by these bow
shocks, as well as the synchrotron peak energies in the forward and
reverse shock systems and the radiation efficiencies at various
frequencies. In this model, the likeliest explanation for the Chandra
emission \citep{M04} is the pulsar A wind just behind the bow shock
caused by the systemic motion in the ISM.  In this case, we predict a
power law spectrum which extends up to $\lesssim 60$ keV.

The eclipse of the pulsar A radio emission near superior conjunction
is best explained as synchrotron absorption by the shocked pulsar A
wind in the bow shock around pulsar B
\citep{Kaspi04,D04,Lyutikov04,Arons04}.  This explanation requires a
relatively large number density of $e^\pm$ pairs, which in turn
requires a relatively low wind Lorentz factor, $\gamma_w\lesssim 100$.
However, the X-ray emission from both of the bow shocks is not very
sensitive to the exact value of $\gamma_w$, and $\gamma_w\sim 10-100$
would only lower the radiative efficiency $\epsilon_{\rm rad}$ and the
X-ray luminosity $L_X$ by a factor of $\sim 2$ (for $p\approx 2$)
compared to $\gamma_w\sim 10^5$.

An alternative explanation for the X-ray emission, is simply emission
from pulsar A \citep{M04,ZH00}.\footnote{Emission from pulsar B is
  unlikely, since $\dot{E}_{\rm rot,B}\sim L_X$, which would require a
  very high efficiency in producing X-rays.} In this case a large part
of the X-ray emission is expected to be pulsed with a period $P_A$.
In contrast, the emission from the bow shock around pulsar B might be
modulated\footnote{Although we find that the emission from the bow
  shock around pulsar B is likely to contribute only a few percent of
  the total X-ray luminosity from this system, it can still produce an
  overall modulation of up to several percent, which might still be
  detectable.} at $P_{\rm orb}$ or $P_B$, while the emission from the
bow shock with the ISM is not expected to be modulated but might be
angularly resolved by Chandra.

\acknowledgments

We thank Roman Rafikov and Peter Goldreich for useful discussions,
and the referee for a careful reading and constructive comments.
JG is supported by the W.M. Keck foundation and by NSF grant PHY-0070928.
PM is supported by the Monell Foundation, NASA NAG5-13286 and NSF AST0098416.

\end{document}